# Enhanced Thermoelectric Properties of 2D Janus Ferromagnetic LaBrI with Strain-induced Valley Degeneracy

Anuja Kumari,[1, a] Raman D. K,[2] Tanu Choudhary,[3] Raju K Biswas,[3, 4] Abhinav Nag,[5] Santanu K. Maiti,[1, b] Jagdish Kumar[2]

[1]*Physics and Applied Mathematics Unit, Indian Statistical Institute, 203 Barrackpore Trunk Road, Kolkata-700 108, India*
[2]*Department of Computational Sciences, Central University of Punjab, Bhatinda-151 401, India*
[3]*Department of Physics, M S Ramaiah University of Applied Sciences, Bengaluru-560 058, India*
[4]*Department of Physics, North Eastern Regional Institute of Science and Technology, Nirjuli, Arunachal Pradesh-791 109, India*
[5]*Department of Physics, GGDSD college, Rajpur, Palampur-176 061, India*

[a]anudhiman45@gmail.com
[b]santanu.maiti@isical.ac.in

Since the successful synthesis of the MoSSe monolayer, which violated the out-of-plane mirror symmetry of TMDs monolayers, considerable and systematic research has been conducted on Janus monolayer materials. By systematically analyzing the LaBrI monolayer, we are able to learn more about the novel Janus material by focusing on the halogen family next to group VIA (S, Se, Te). The structural optimizations have been carried out using pseudopotential based Quantum espresso code. Computed structural parameters are in good agreement with literature reports. The optimized crystal structures were used for computing effect of strain on electronic and thermoelectric properties. Dynamical stability predicts that this material can withstand up to 10% of tensile strain. Computed electronic structure reveals material to be indirect wide band gap ferromagnet with magnetic moment $1\mu_B$. With increase in the biaxial tensile strain the band gap decreases. Furthermore, the computed magneto-thermoelectric properties predict high Seebeck coefficient ~ 400 μV/K and low thermal conductivity of ~ 0.93 W/m.K in LaBrI which results in a high ZT ~ 1.84 for 4% strain at 800 K. The present study supports the fact that tensile strain on ferromagnetic LaBrI material can further enhance TE properties making it to be a promising material for TE applications at higher temperatures.

## 1. Introduction

Clean energy sources and renewable energy technology are the utmost priority of scientists, ecologists and researchers to combat the global warming and climate change [1]-[5]. Among various factors, waste heat is one of the prime concerns causing it. By scavenging wasted heat, we can save energy and promote efficient energy production [6]-[11]. Thus, it is significant to explore new ways to generate, convert, and store energy. Thermoelectric generators (TEGs) are a highly effective solution in managing waste heat as they can convert waste thermal heat into useful electrical energy [12]-[16]. TEGs are significantly known for their applications in deep space exploration because of immobile parts, silent working, and steadiness during years of operation. A thermoelectric module is simply composed of parallel thermal and series electrical connections between pairs of highly doped p-type and n-type semiconductors. For any good TE module, these p-type and n-type TE materials are very crucial. Thus, researchers have explored many different TE materials for power generation at various temperatures. Although conversion efficiency is not yet achieved for commercialization of this technology and is yet a challenge [17]-[20]. The efficiency of any thermoelectric material is assessed by a dimensionless parameter called a figure of merit denoted by *ZT* as [21],[22]

$$ZT = \frac{S^2 \sigma T}{\kappa_e + \kappa_l}. \tag{1.1}$$

In the above equation, *S* represents the Seebeck coefficient, σ represents the electrical conductivity, $\kappa_e$ represents electronic and $\kappa_l$ represents phononic thermal conductivity. Thus, a factor known as the power factor, $S^2\sigma$, indicates the electric performance of the material. To achieve optimal thermal performance at a certain temperature, a higher power factor and lower thermal conductivity are desirable [23]-[26]. However, for any given material three parameters *S*, *σ*, and $\kappa_e$ are coupled together, and optimizing them simultaneously is a daunting task [27]-[32]. This unveils various strategies to improve TE materials such as power factor augmentation by carrier filtering [33]-[37], carrier pocket engineering [38]-[40], complex structures [41]-[42], low dimensional structures [43]-[45] and valley degeneracy[46] etc., that improve *ZT* significantly. Also, improved phonon scattering using nano-structuring has been used to raise the TE material's energy conversion efficiency [47].

With the prediction of excellent transport properties and large power factors of graphene [48] and other materials [49]-[52], many layered materials have been investigated for thermoelectric applications. Due to their indifferent structural, physical, and chemical properties and easy synthesis using a variety of methods such as chemical vapor deposition (CVD), physical vapor deposition, mechanical exfoliation etc., they are of recent interest in thermoelectric [53]-[56]. Gu and Yang [57] predicted that altering the stoichiometry of the structure can lead to a significant reduction in the lattice thermal conductivity. The *ZT* can be efficiently increased by lowering the lattice thermal conductivity. Recently, the synthesis of Janus MoSSe monolayer [58] inspired the research community to explore a new family of 2D materials named Janus materials. Due to highly asymmetric structural characteristics, Janus materials exhibit diverse properties, including finite out-of-plane dipoles, giant Rashba splitting, a wide range of direct and indirect band gaps, etc [58]. In Janus transition metal dichalcogenides (TMDCs) monolayer, the layer of metal atoms is sandwiched between two layers of dissimilar chalcogen atoms. They possess an extra degree of freedom to tune material properties while retaining the exotic characteristics of their parent structure. Numerous works in the literature suggested that these Janus materials can be a potential material for TE applications [52]. Furthermore, recent theoretical studies suggest that the Janus monolayer can be useful in a variety of emerging fields [59]. For thermoelectric application, study by Abhishek *et al.*[60] reported a high *ZT* of 2.56 for the WSTe Janus monolayer at higher temperature, another Janus monolayer ZrSSe studied by S. Zhao *et al.* [61] provides a high *ZT* of 4.41 and 4.88 for p-type and n-type materials with a tensile strain of 6%. Similarly, *ZT* of 2.54 for PtSeTe Janus monolayer is observed at 900K by Wang *et al.* [62] and *ZT* ~ 3.00 is reported by Shivani *et al.*[63] in MoSSe Janus monolayer at 1200K.

Magnetic Janus materials have also been proposed recently such as manganese chalcogenides [64] and chromium trihalides [65]. Jiang and Kang predicted a 2D magnetic Janus ferrovalley material LaBrI to be a stable ferromagnetic electrode [66] by employing the first-principles investigations. They proposed that the intrinsic magnetism and asymmetric nature of these materials result in a large valley degeneracy and hence making these materials a promising choice for various applications. Interestingly, due to valley degeneracy, the WSSe Janus monolayer shows good thermoelectric properties [67]. Motivated by their idea, we have investigated various properties of 2D magnetic Janus material LaBrI for thermoelectric applications. Janus monolayer of LaBrI has broken out-of-plane symmetry from $D_{3h}$ ($LaBr_2$) to $C_{3V}$ (LaBrI) with space group *P3m1* (No. 156) as shown in Figure 1. This monolayer is a three-atomic thick layered structure with La atoms sandwiched between two halogen atoms Br and I having reflection asymmetry on the mirror plane.

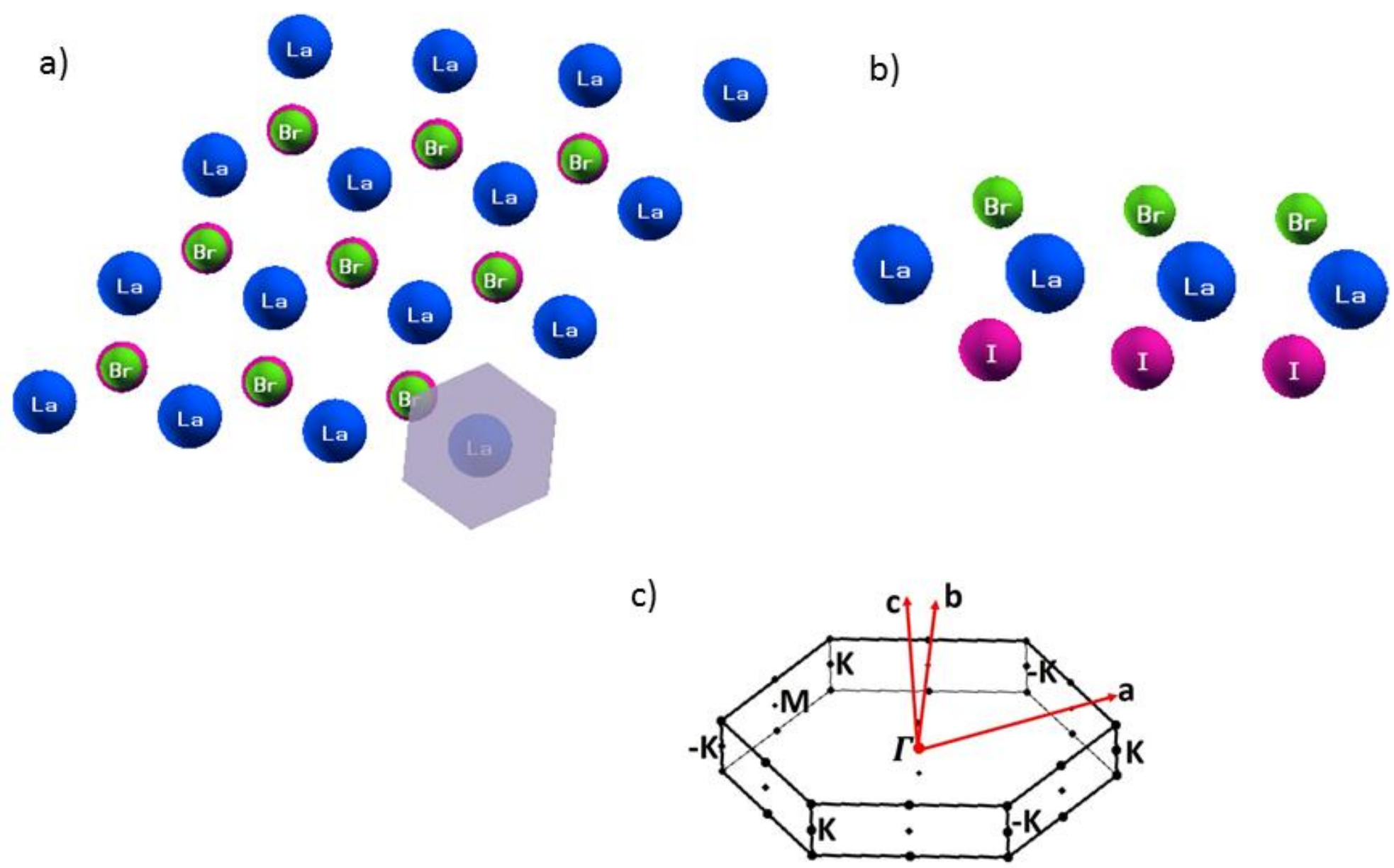

Figure 1: Crystal structure of 2D Janus monolayer of LaBrI (*P3m1*: no. 156) a) top view b) side view and c) Brillouin zone with high symmetry points.

In the present study, we have systematically investigated the structural properties, dynamical stability, electronic, and thermoelectric properties of this material. To the best of our knowledge, *a study on the thermoelectric properties of this material was not reported yet*. Furthermore, we present modulated ferrovalley for up-spin and down-spin-charge carriers in electronic band structure which enhances with increase in strain. Due to this modulation in valley degeneracy, low thermal conductivity is observed which enhances the figure of merit *ZT* to be more than one. Thermoelectric coefficients as a function of chemical potential at various temperatures predict *ZT* >1 for p-type and n-type dopings at different temperatures. Obtained low thermal conductivity and high *ZT* values tuned with strain are the essential findings of our study that predicts this material to be an efficient thermoelectric material.

The present paper is arranged as follows. In section 2 we discuss the theoretical formalism of density functional theory (DFT) used in computing all the calculations. In section 3 we further present and discuss lattice parameters, dynamical stability based on phonon dispersion curves and various thermoelectric parameters such as the Seebeck coefficient, electrical conductivity, power factor, thermal conductivity and *ZT* of the material. Finally, in section 4, we conclude with significant outcomes of present study.

## 2. Computational Details

All the calculations were carried out in the framework of the density functional theory (DFT). Janus monolayer of LaBrI is designed from pristine $LaBr_2$ monolayer by replacing Br atoms with Iodine atoms. With the optimized lattice parameters of LaBrI, the calculations of electronic and dynamical properties have been carried out by employing pseudopotential based Quantum Espresso package [68]. The exchange and correlation interaction among the electrons are described using the non-empirical *generalized gradient approximation* within the framework of PBE (Perdew-Burke-Ernzerhof) formalism which performs reasonably accurate for a wide range of materials [69]. The charge density and wave function cutoff were set to 320 Ry and 40 Ry, respectively, throughout the calculations. All calculations were performed using norm-conserving scalar relativistic and fully relativistic pseudo-potentials. Due to the presence of heavy elements, spin orbit coupling (SOC) effect was incorporated using fully relativistic norm-conserving pseudopotentials for all the three atoms. We have used a Monkhorst-Pack of 12×12×2 *k*-points for electronic properties and 36×36×2 for thermoelectric properties for sampling the Brillouin zone (BZ). The lattice parameters and atomic positions were obtained by relaxing the structures to force tolerance of $10^{-4}$ Ry/Bohr. A large vacuum space of 15 Å is used along the z-axis to prevent any artificial interactions between periodic images of slabs along the vertical direction. The phonon dispersion calculations were employed for the accessing dynamical stability. We obtain the phonon dispersion with a *q* mesh of 3×3×1 with a strict convergence threshold of $10^{-10}$ Ry in the self-consistent calculation using density functional perturbation theory (DFPT). The thermoelectric coefficients were computed using the semi-classical Boltzmann transport theory within constant scattering time approximation (CSTA) and the rigid band approximations, as implemented in the BoltzTrap code [70]. To compute the electronic part of thermal conductivity, the constant value of relaxation time is utilized in BoltzTrap code.

According to semi-classical Boltzmann transport theory, once we get the first-principles band structure calculations we can define *Energy projected conductivity tensor '$\Theta(\varepsilon)$'* in terms of conductivity tensor as [70],

$$\Theta(\varepsilon) = \frac{1}{N}\sum_k \sigma(k)\left( - \frac{\delta(\varepsilon - \varepsilon_k)}{d\varepsilon} \right) \quad . \tag{2.1}$$

It shows the distribution in energy and represents the contribution of electrons having a specific energy '$\varepsilon$' and '$N$' represents the number of $k$-points sampled in the Brillouin zone. For a unit cell crystal having volume $\Omega$, the transport coefficients can be evaluated by integrating this conductivity distribution as;

$$\sigma(T,\mu) = \frac{1}{\Omega}\int \Theta(\varepsilon)\left[-\frac{\delta f_{\mu}(T,\varepsilon)}{\delta k}\right]d\varepsilon \qquad 2.2$$

$$\kappa_{e}(T,\mu) = \frac{1}{eT\Omega}\int \Theta(\varepsilon)(\varepsilon-\mu)^{2}\left[-\frac{\delta f_{\mu}(T,\varepsilon)}{\delta k}\right]d\varepsilon \qquad 2.3$$

$$S(T,\mu) = \frac{1}{eT\Omega T(T,\mu)}\int \Theta(\varepsilon)(\varepsilon-\mu)\left[-\frac{\delta f_{\mu}(T,\varepsilon)}{\delta k}\right]d\varepsilon \qquad 2.4$$

where, $T$ is the equilibrium temperature, $\mu$ is the chemical potential, $\kappa_e$ is the electronic part of the thermal conductivity and $S$ refers to the Seebeck coefficient.

We have also computed lattice thermal conductivity $\kappa_l$ by solving Boltzmann Transport Equation (BTE) under Relaxation Time Approximation (RTA). At thermal equilibrium, the phononic distribution obeys Bose-Einstein statistics [71][72]. In the steady-state, the phonon distribution function is expressed as,

$$\frac{df_{\lambda}}{dt} = \frac{df_{\lambda}}{dt}\,|_{diffusion} + \frac{df_{\lambda}}{dt}\,|_{scattering} = 0 \qquad 2.5$$

where $f$, t, λ are the distribution function, time, and phonon mode respectively. The phononic distribution $f_\lambda$ of two- and three-phonon scattering process is modelled under RTA where third order anharmonic force constants are computed to obtain the three-phonon scattering rate. The final lattice thermal conductivity tensor is expressed as,

$$k_{\alpha\beta} = \frac{1}{k_B T^2 \Omega N}\sum_{\lambda} f_0(f_0+1)(\hbar\omega_{\lambda})^2 v_{\lambda}^{\alpha} F_{\lambda}^{\beta} \qquad 2.6$$

where Ω, $f_0$, $v_\lambda$, and N are the volume of the unit cell, equilibrium Bose-Einstein distribution function, group velocities of phonon mode and the number of $q$ points in the Brillouin zone respectively.

The lattice thermal conductivity calculations were performed using Phono3py code [73]. The second and third order force constants were calculated by using 2x2x1 supercells, which required force calculations for 333 number of crystal structures containing 12 atoms. Quantum Espresso code was utilized to perform static calculations. The *k*-grid size was set to 5x5x1. The kinetic energy cutoff for wave-functions was set to 50 Ry and the force and energy threshold were set to $10^{-5}$ Ry/Bohr and $10^{-6}$ Ry using Phono3py. A well converged *q*-mesh size of 20x20x1 as shown in Figure S1 (see Supplemental Material [74]), was chosen for calculating thermal conductivity tensors under BTE-RTA (relaxation time approximation). The calculated lattice thermal conductivity results were then scaled to make it independent of the vacuum thickness. The values were multiplied by the factor of L/*d*, where L is the length of the z-axis and *d* is the sum of layer thickness and the van der Waals radii of the outermost atoms.

Finally, the figure of merit (*ZT*) can be obtained by,

$$ZT = \frac{S^2 \sigma T}{\kappa_e + \kappa_l}. \tag{2.7}$$

Here, *ZT* is related to thermal to electrical energy conversion efficiency of any material. To obtain more accurate value of *ZT*, phononic thermal conductivity plays a significant role and literature suggests that strain mediated valley degeneracy reduces thermal conductivity significantly [67]. Thus, systematic study of TE properties of Janus monolayer is presented in this manuscript.

.

# 3. Numerical Results and Discussion

## 3.1 *Structural properties*

By allowing the crystal to relax to a force tolerance of $10^{-4}$ Ry/Bohr, the equilibrium lattice parameters for all the compositions under investigation have been determined. Using the GGA-PBE as the exchange-correlation, we have calculated the lattice constants and other related parameters which are listed in Table I. The values agree well with the previous reports in literature [66].

Table I: Variation in bandgap with different strain on Janus LaBrI system obtained using GGA-PBE exchange and correlation. Computed energy splitting due to spin-orbit coupling at *K* and –*K* high symmetry points.

| S. No. | Biaxial tensile strain | Parameter | LaBrI | Energy splitting (meV) |
|---|---|---|---|---|
| 1 | 0% | Lattice constant (Å) | 4.24 (4.24)[66] | 28 |
| | | Bandgap (eV) | 2.44eV (↑spin)<br>1.16eV (↓spin) | |
| 2 | 4% | Bandgap (eV) | 2.75eV (↑spin)<br>0.45eV (↓spin) | 49 |
| 3 | 8% | Bandgap (eV) | 3.17eV (↑spin)<br>0.14eV (↓spin) | 61 |
| 4 | 10% | Bandgap (eV) | 3.20eV (↑spin)<br>0.08eV (↓spin) | 69 |

### *3.2 Lattice Dynamics and Dynamical Stability*

The thermoelectric performance of a material drastically changes with the implication of strain (ε). The tensile strain is defined as $\varepsilon\,\% = \left(\frac{a-a_0}{a_0}\right) \times 100$ where $a_0$ is the equilibrium relaxed lattice constant. We examined the effect of biaxial tensile strain on the dynamical stability of LaBrI by calculating phonon dispersion curves at the steps of Δε=4%. Figure 2, illustrates the phonon dispersions at various strains for LaBrI Janus monolayers along different high-symmetry directions.

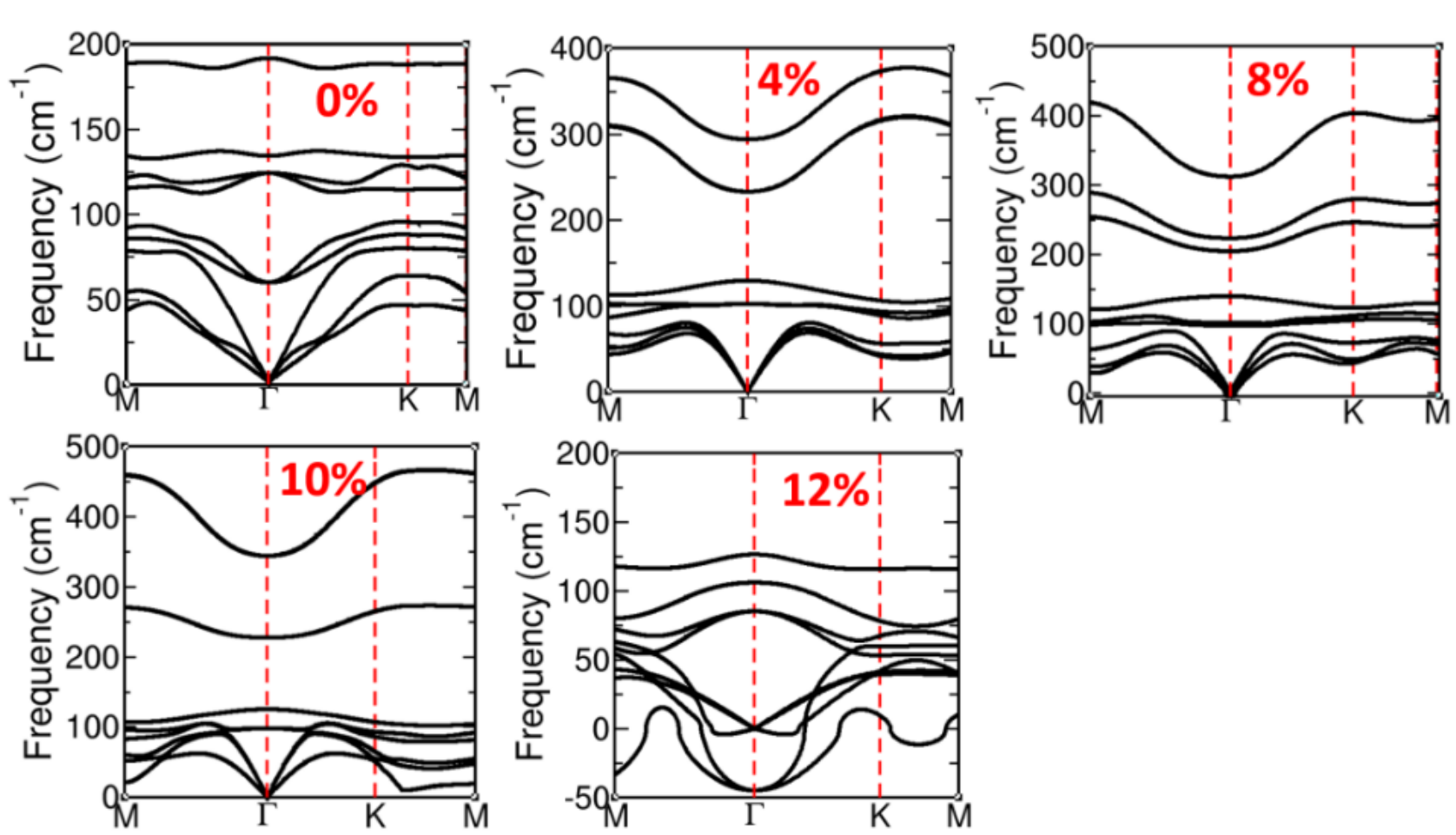

Figure 2: Phonon dispersion for LaBrI at different biaxial tensile strain along high symmetry *k*-points.

The computed phonon dispersion for LaBrI comprises three acoustic and six optical branches in the dispersion curves which is due to the one La atom, one Br atom, and one I atom in the unit cell. The phonon dispersion curves relate the dependence of phonon frequencies for all branches along selected high symmetry path (see supplemental material [74]). The dynamical stability of a material is associated with the real and imaginary frequencies of the phonon dispersion curves. Any imaginary frequency in the phonon dispersion curve signifies the non-restorative force which results in the lowering of potential energy of the atoms when displaced from equilibrium positions indicating dynamical instability of the structure. Computed phonon dispersion curves for LaBrI do not contain any imaginary phonon frequencies up to 10%. This indicates that LaBrI monolayer is dynamically stable under tensile strain up to 10%. Further increase in the strain led to imaginary frequencies in the phonon dispersion curve and hence material gets unstable beyond this limit.

### *3.3 Electronic Properties*

The spin polarized electronic band structures for the studied monolayer under various strain values are shown in Figure 3. Band structures highlight that GGA-PBE predicts LaBrI to be a semiconductor with an indirect band gap of 0.51 eV. Due to spin polarization, VBM and CBM near the Fermi level possess opposite spin that makes this monolayer a spin semiconductor. It can be seen from Figure 3 that near Fermi level the bands have local minima for up spin and down spin which are significant and referred to as valleys. These valleys offer additional degree of freedom for the electron along with its charge and spin. With an increase in strain, the electronic valley degree of freedom can be manipulated and hence it can be used as to tune the energies of these valleys. LaBrI is a ferromagnetic material with a magnetic moment per unit cell of ~ $1\mu_B$ [66]. The magnetism is localized in the interstitial region of La atoms. Most contribution of magnetic moment comes from La atoms while it is negligible at Br and I atomic sites.

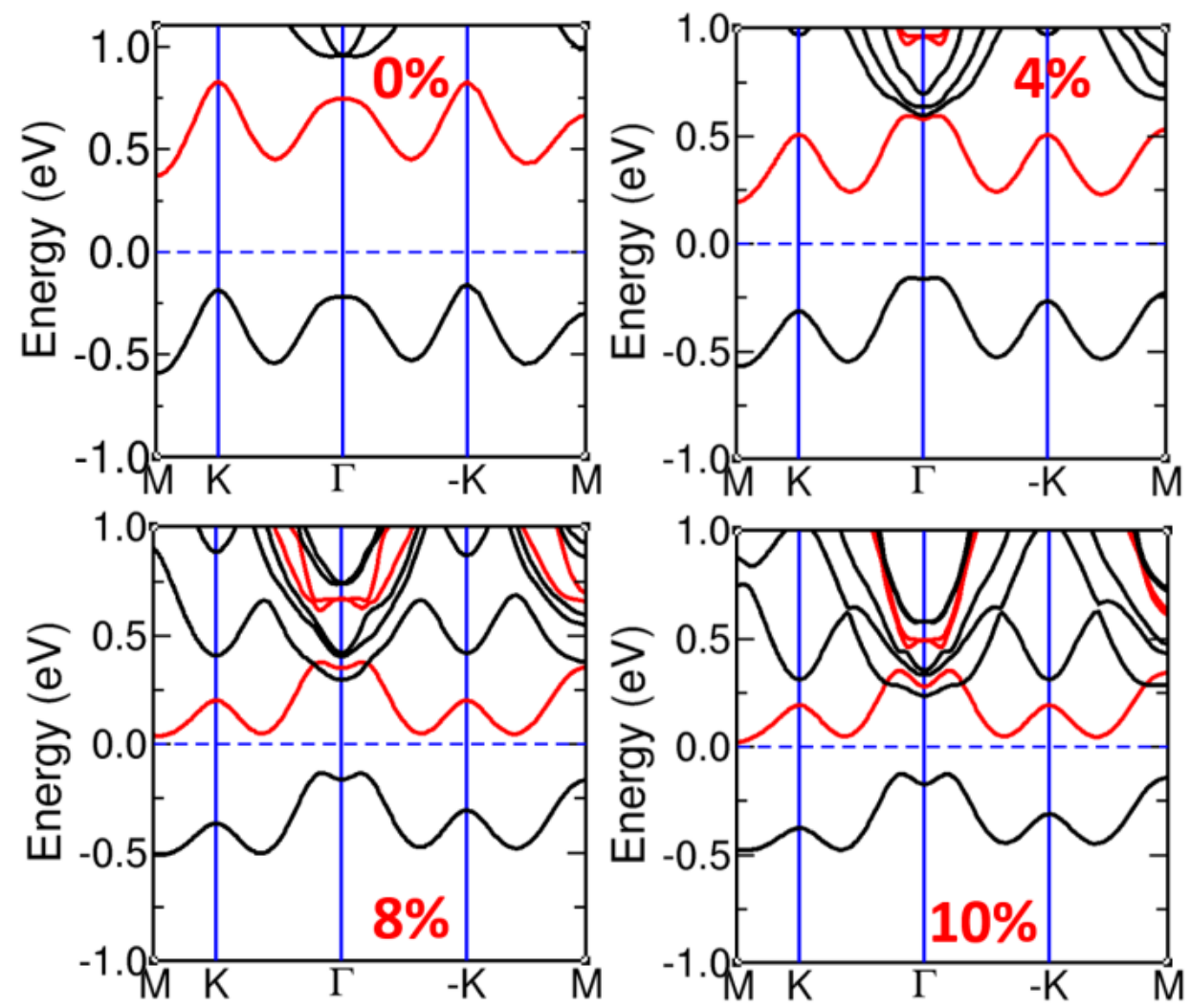

Figure 3**:** Electronic band structure showing up spin (black color) and down spin (red color) obtained using GGA-PBE functional for various strains on LaBrI

It is apparent from Figure 3, that LaBrI shows an indirect band gap from *Γ* to M point in the Brillouin zone. The striking feature in the band structure is the presence of valley polarization at *K* and –*K* high symmetry points. In Janus materials the broken symmetry of the crystal, leads to formation of such valleys which create additional degree of freedom for tuning the electronic properties. With increase in strain these valleys can be polarized further as they shift closer to the Fermi level. We observed significant changes in the conduction band region when a tensile strain is applied to the monolayer. The states at both the *Γ* and *M* points increase as the strain increases and the states at *Γ* point shifts at slower rate than those at *M* point. Moreover, with the increase in tensile strain the band gap decreases. Such decrease in the bandgap can be related to the increase bond length with tensile strain leading to reduction in the coupling and energy band splitting. We found that when a biaxial strain is applied, the band valleys in the Janus monolayer can significantly be tuned. Such tuning of bands due to strain has a significant effect on the band curvatures and hence the effective mass of the charge carriers in valance and conduction bands and hence the transport properties.

The studied Janus layer has heavy atoms such as La and I which will have significant contribution of the spin-orbit coupling on the band structure. Therefore, we have also investigated the effect of SOC+GGA on electronic and transport properties. With the application of SOC there is no significant change observed in the band structure for both unstrained and strained systems. For strained system, the SOC exhibits symmetric band curvatures resulting in insignificant effect on the transport properties of these monolayers. The inclusion of SOC with an initial splitting in the magnetic moment led to the energy splitting which depends upon the direction of the magnetization. Due to the magnetization along the –z axis, obtained energy band maximum at –*K* is 28 meV higher than energy band maximum at *K* point for unstrained system as shown in Figure 4.

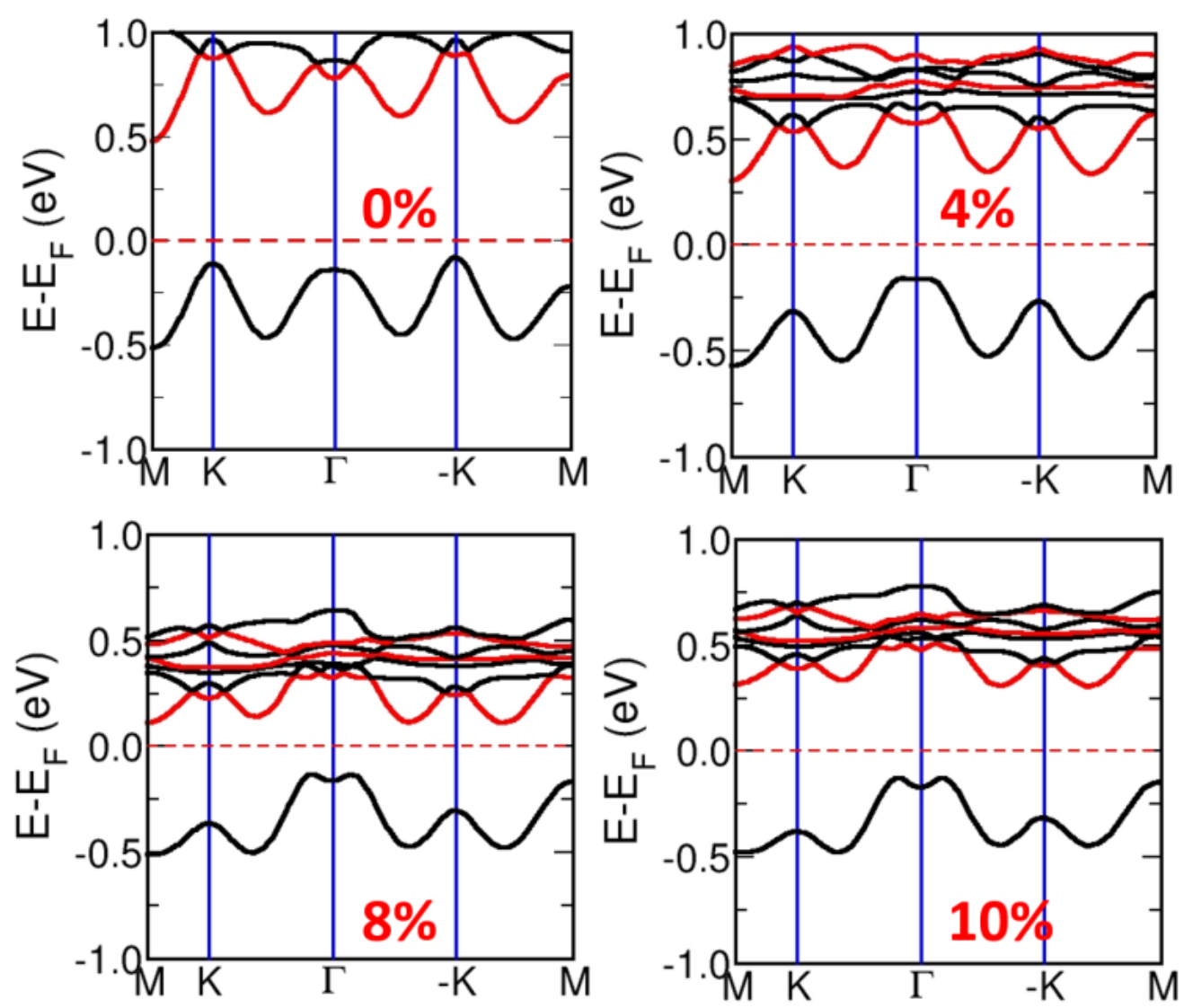

Figure 4: Electronic band structure showing up spin (black color) and down spin (red color) obtained using GGA+SOC for different strain.

Without SOC these energy bands at *K* and -*K* are energetically degenerate showing valley polarization. Further, increase in strain with SOC led to increase in the energy splitting as tabulated in Table I.

Since, the SOC is significantly higher in heavier systems as compared to lighter systems which lead to the strong band splitting in VBM and CBM resulting in decreasing bandgap with strain. Such strong SOC in Janus monolayers and tunable Fermi-level with strain or applied field can be a good strategy for tailoring these materials for valleytronics and spintronics applications.

### *3.4 Thermoelectric Properties*

The thermoelectric coefficients of the considered monolayer have been computed which include the Seebeck coefficient (*S*), electrical conductivity (σ), electronic part of thermal conductivity ($\kappa_e$) and power factor ($S^2\sigma$) by solving the Boltzmann transport equations within constant scattering time approximation (CSTA) and rigid band assumption. We have investigated the impact of doping the charge carriers on the transport coefficients by varying the chemical potential ($\mu$). The transport properties are dependent significantly on the location of the chemical potential. The negative and positive doping levels of the chemical potential refer to hole (*p*-type) and electron (*n*-type) dopings, respectively.

### *3.4.1 Seebeck coefficient (S)*

The Seebeck coefficient is the induced electric potential due to thermal gradient across a thermocouple. Therefore, a higher Seebeck coefficient signifies higher electric potential across thermal gradient and is essential for higher $ZT$ or good thermoelectric material. For a given material, the Seebeck coefficient is related to carrier concentration '$n$' as [72]

$$S = \frac{8\pi^2 k_B^2 T}{3eh^2} m^* \left(\frac{\pi}{3n}\right)^{\frac{2}{3}} \quad 3.1$$

where, $T$ is the temperature, $k_B$ is Boltzmann's constant, $e$ is electronic charge, $m^*$ is effective mass, and $h$ is Planck constant. Within constant scattering time approximation (CSTA) the electronic conductivity tensor $\Theta(\varepsilon)$ using equation 2.1 as discussed in section 2, can be obtained only from the velocities that follow from electronic band structure. The chemical potential $\mu$ is used as a variable parameter to study the effect of doping. The obtained Seebeck coefficient for the monolayer of LaBrI as a function of carrier concentration is depicted in Figure 5. Seebeck coefficient is directly proportional to the effective mass of the carrier and inversely proportional to the carrier concentration near the Fermi level. For unstrained monolayer, the highest Seebeck coefficient of ~ 466 μV/K is obtained at 600 K for up spin-charge carriers, which is evident from the band curvatures shown in Figure 5 as near the Fermi level heavy holes with higher effective mass and low carrier density led to high Seebeck coefficient. Although for down spin-charge carriers, temperature does not have a significant effect and a coefficient of ~112μV/K was recorded with positive doping levels. The Seebeck coefficients for different biaxial tensile strains at various temperatures 600K, 800K, and 1200K are also illustrated in these plots. For spin-up charge carriers Seebeck coefficient decreases to ~330μV/K with strain due to contribution of light holes which have low effective mass and higher carrier concentration. Similar trends are observed with an increase in temperature. However, for down spin charge carriers, it increases to ~480μV/K with positive doping levels at 600K. At higher temperatures 800K and 1200K, we get a significantly low value of the Seebeck coefficients.

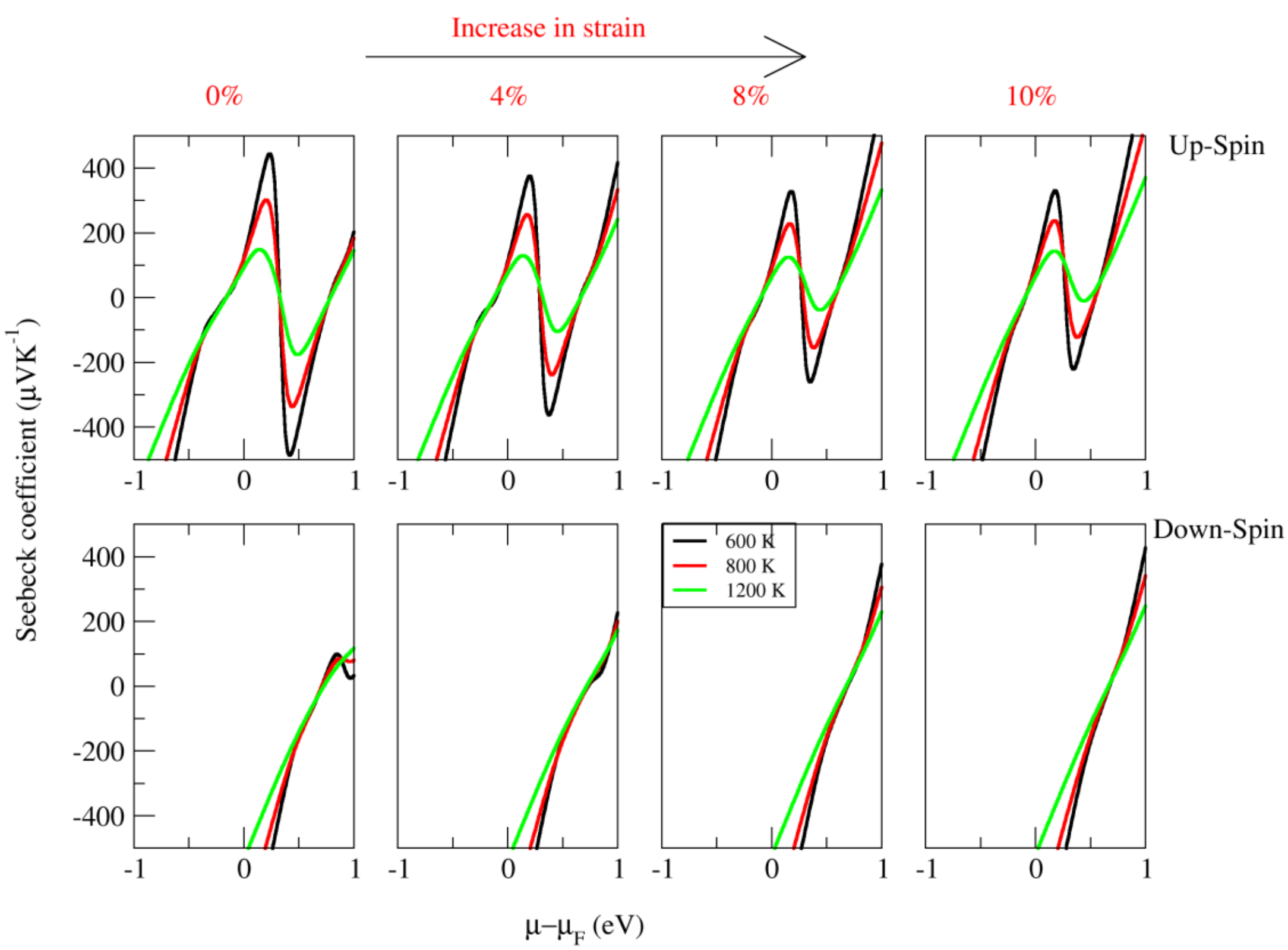

Figure 5: Seebeck coefficient as a function of chemical potential at 600K, 800K, and 1200K for spin-up and spin-down charge carriers.

### *3.4.2 Power Factor ($S^2\sigma$)*

The *ZT* of any TE material is directly proportional to its power factor which is defined as $PF = S^2\sigma$. It significantly comprises of the electronic conductivity and Seebeck coefficient.

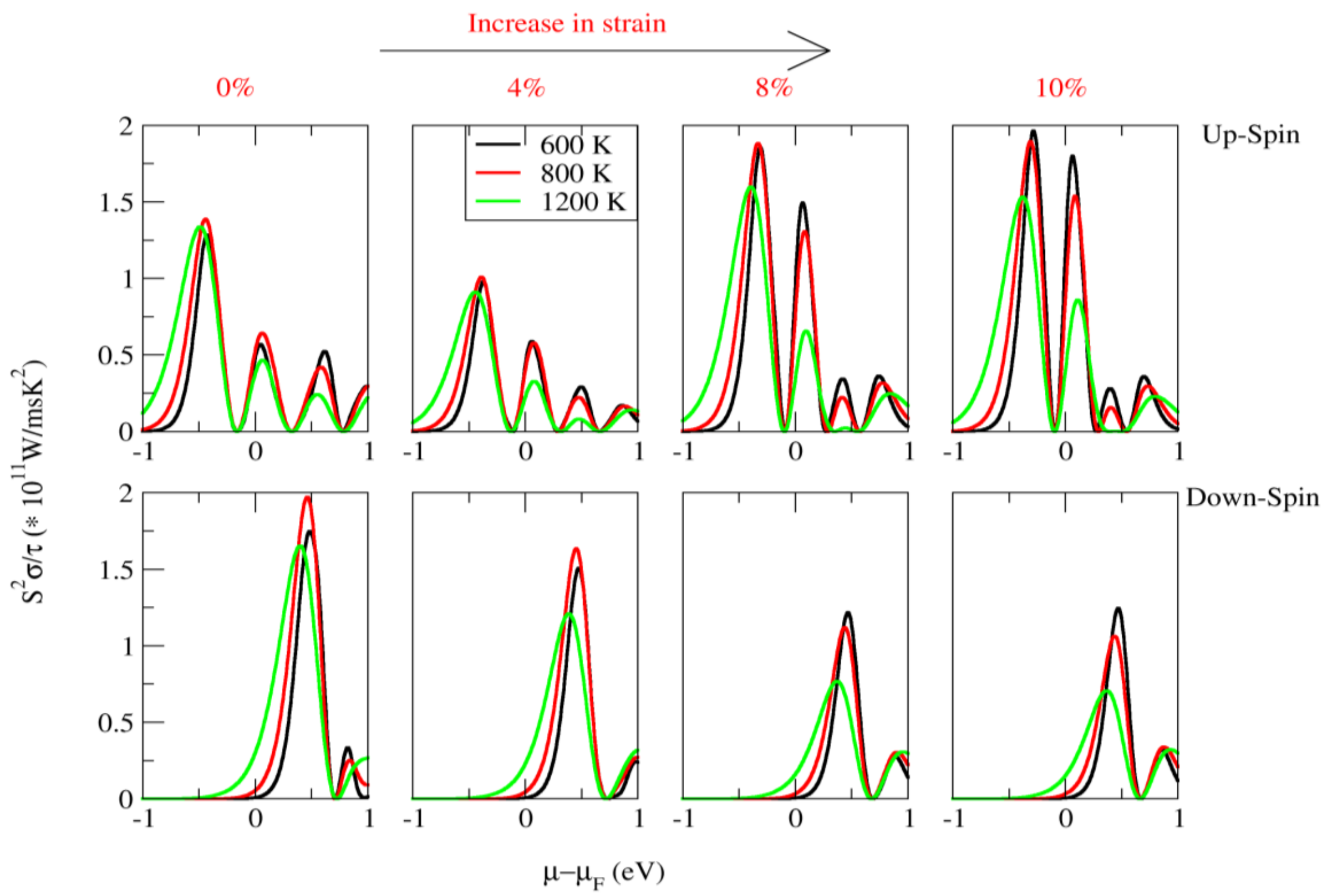

Figure 6: Spin resolved power factor as a function of chemical potential at 600K, 800K, and 1200K.

Figure 6 illustrates the calculated power factor as a function of chemical potential employed at different strain under various temperatures. For unstrained LaBrI, spin up charge carriers show increase in PF from ~$1.28 \times 10^{11} W/m.K$ at 600K to ~$1.34 \times 10^{11} W/m.K$ at 1200K and spin down charge carriers shows increase in PF from ~$1.78 \times 10^{11} W/m.K$ at 600K to ~$1.96 \times 10^{11} W/m.K$ at 800K. Further, at 800K with 8% strain, spin up carriers shows PF ~$1.83 \times 10^{11} W/m.K$ with p-type doping and spin down charge carriers show lower PF ~$1.04 \times 10^{11} W/m.K$ with n-type doping. At high temperatures, we can observe decrease in the power factor. Thus, strain significantly modifies the band curvatures yielding changes in effective mass of the charge carriers ultimately leading to changes in power factor. High band curvatures for up spin near Fermi level shows light holes with less effective mass which result in high power factor for unstrained system. With increase in strain band curvatures decreases resulting in heavy holes with high effective masses and hence the relevant effect can be seen in the power factor for up spin and down spin charge carriers.

### 3.4.3 *Thermal conductivity*

Thermal conductivity of a material is a sum of electronic thermal conductivity ($\kappa_e$) and lattice thermal conductivity ($\kappa_l$). Solving Boltzmann transport equations we have obtained electronic thermal conductivity $\kappa_e$. To obtain an optimization between electronic conductivity ($\sigma$) and electronic thermal conductivity ($\kappa_e$), we need to understand the relation among them. At an absolute temperature, they are related by Wiedemann-Franz law as $\frac{\kappa_e}{\sigma} = LT$ [75][76], where ($\kappa_e$) is the electronic thermal conductivity, ($\sigma$) is the electrical conductivity, $L$ is the Lorenz number and $T$ is the absolute temperature. According to this law that refers to the simplest free electron model, the electrical and thermal conductivities should be in direct relation to get the Lorenz number as a constant at a particular temperature. However, in a more realistic scenario of complex interactions within the crystal structure, we can find a way to optimize of thermoelectric performance of any material.

Figure 7 depicts the electronic thermal conductivity of this monolayer for *n*- and *p*-type dopings at different temperatures. It can be observed from the plots that in an unstrained system, the spin-up and spin-down charge carriers have a lower electronic thermal conductivity of ~$1.14\ W/m.K$ and ~$1.02\ W/m.K$ at 600K respectively. However, the increase in strain has prominent changes in both spin up and spin down charge carriers as with 4% strain, the thermal conductivity decreases to ~$0.62\ W/m.K$ for spin up and it increases to ~$1.64\ W/m.K$ for spin down charge carriers at 600K. A similar trend is observed concerning increase in temperature respectively. With increase in strain to 4%, electronic thermal conductivity decreases for spin up and increases for spin down charge carriers. Therefore, overall the tensile strain seems to be favorable for enhancing the thermoelectric performance of this monolayer.

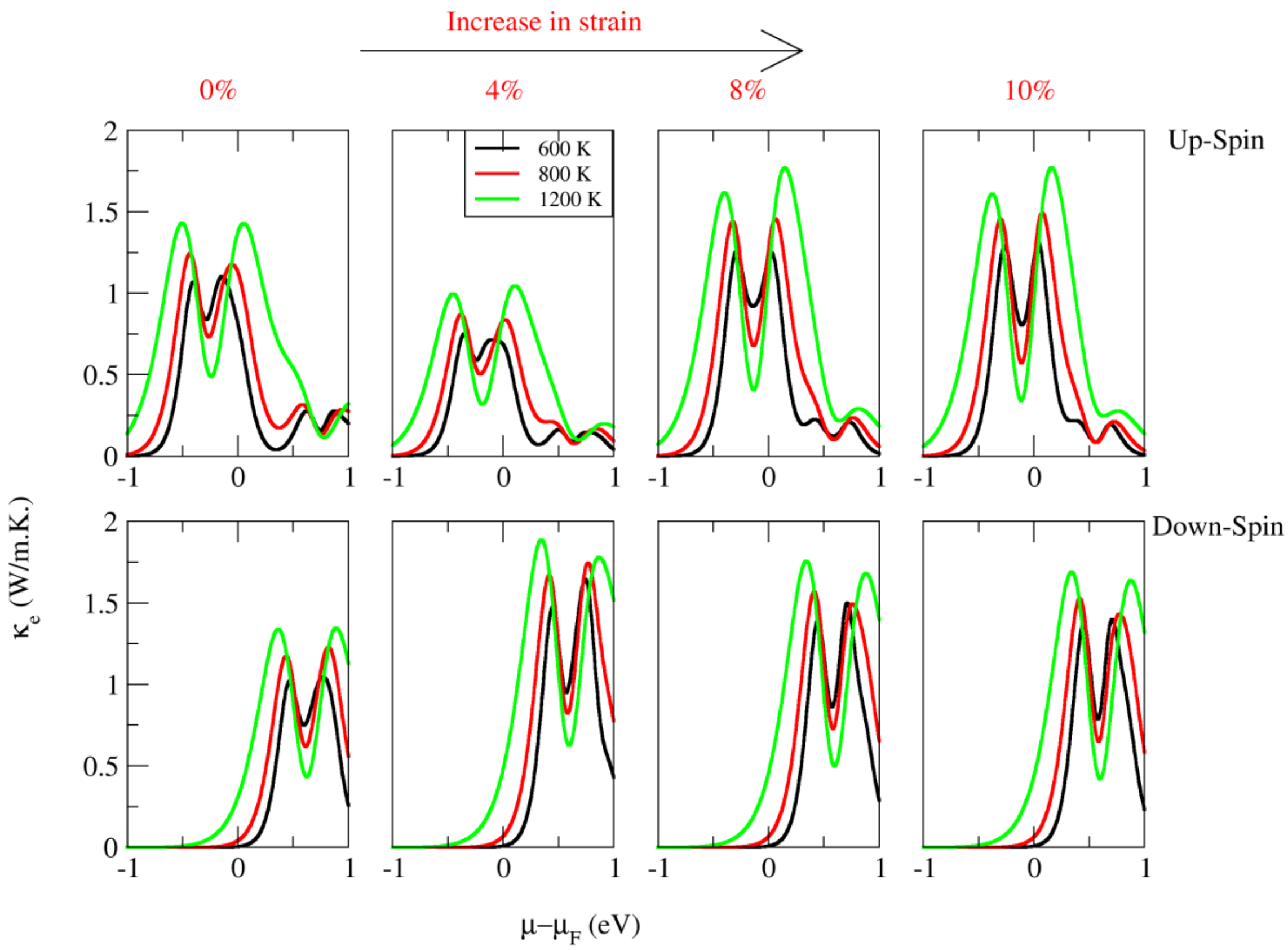

Figure 7**:** Variation of electronic part of thermal conductivity with chemical potential at various temperature for up spin and down spin charge carriers. Here, $\kappa_e$ is scaled using relaxation time $\tau = 10^{-14}$ sec and for $\tau = 10^{-13}$ sec the values will be scaled by multiplying a factor of 10.

For good efficiency or for a high figure of merit, lattice thermal conductivity ($\kappa_l$) must be low. At 300K, lattice thermal conductivity of unstrained LaBrI monolayer was found to be ~5.245 W/m.K which further decreases to 0.059 W/m.K with 10% strain. Interestingly, computed lattice thermal conductivity at higher temperature is significantly low (< 1.0 W/m.K) for LaBrI as presented in Table II.

Table II: Lattice thermal conductivity computed using phono3py at 600K and 800K.

| **S.No** | **Strain %** | **Thermal conductivity (W/mK)** | |
|---|---|---|---|
| | | **600K** | **800K** |
| 1 | 0% | 2.604 | 1.951 |
| 2 | 4% | 1.241 | 0.930 |
| 3 | 8% | 0.428 | 0.322 |
| 4 | 10% | 0.029 | 0.021 |

As depicted from Figure 8, for unstrained system, with the increase in temperature from 600K to 800K, thermal conductivity decreases from ~2.604 W/m.K to ~1.951W/m.K. Although application of 4% strain slightly decreases the $k_l$ to ~0.930 W/m.K at 800K. We noticed a decrease in the lattice thermal conductivity with steady increase in the tensile strain. The decrease in lattice thermal conductivity may be related to increased anharmonic interactions with increase in volume, leading to frequent scattering [77][78]. Therefore, we can infer that low thermal conductivity at higher temperature might result to higher *ZT* of this monolayer.

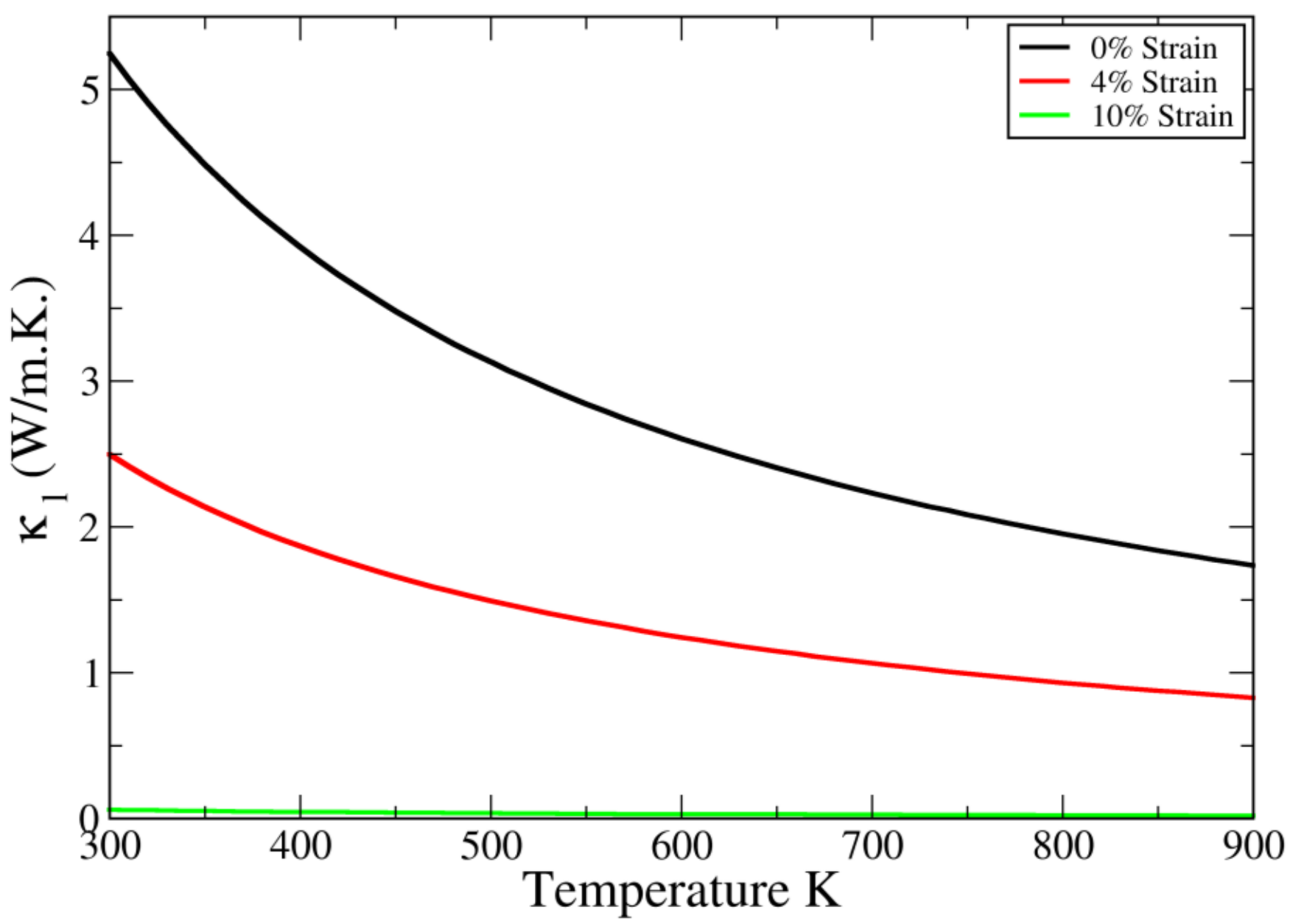

Figure 8: Variation of phononic part of thermal conductivity with temperature at different strain.

### *3.4.4 Thermoelectric Figure of Merit (ZT)*

We further computed the dimensionless figure of merit to predict the thermoelectric efficiency by employing above discussed transport coefficients. Since the electronic transport coefficients are computed within *constant relaxation time approximation,* we have considered two constant values of relaxation to further predict the *ZT*. Considering constant value of $\tau = 10^{-13}$ sec and $10^{-14}$ sec we have computed *ZT* as shown in Figure 9. Computed results with $\tau = 10^{-14}$ sec for unstrained LaBrI have *ZT* < 1 whereas $\tau = 10^{-13}$ sec results ZT = 1.24 at 800K. Variations of temperature and strain have significant effect on *ZT*. For $\tau = 10^{-13}$ sec, *ZT* enhanced from ~0.85 for unstrained LaBrI to ~1.27 with 4% strain for down spin with n-type doping at 600K. At 800K, with 4% strain we obtained highest *ZT* of ~1.84 for down spin with n-type doping. Computed *ZT* with higher strain has slowly decreased *ZT* of 1.74 for 8% strain and 1.80 for 10% strain. Similarly, for $\tau = 10^{-14}$ sec the *ZT* value at 800K is 1.72 and at 600K becomes 1.68 with 10% strain. Thus, *ZT* calculated with $\tau = 10^{-13}$ sec clearly signifies higher *ZT* with 4% strain which signifies strain enhanced thermoelectric figure of merit.

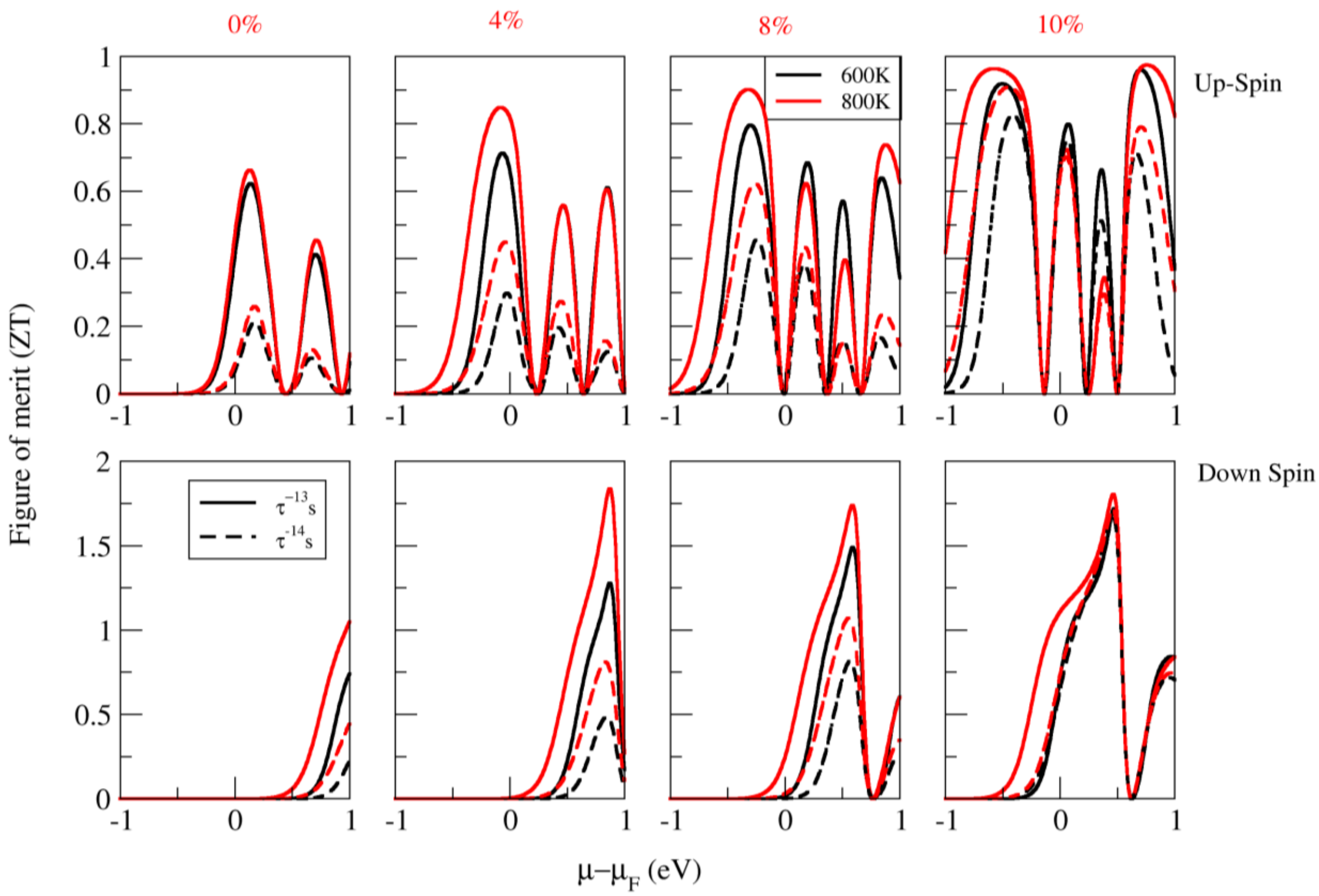

Figure 9: *ZT* as a function of chemical potential in different temperature regions with two values of relaxation times solid lines representing $\tau = 10^{-13}$ sec and dashed lines representing $\tau = 10^{-14}$ sec .

Moreover, the studied Janus monolayer LaBrI exhibits higher *ZT* values which are prominent for potential thermoelectric materials.

## 4. Conclusions

We have systematically investigated the structural parameters of 2D Janus ferrovalley LaBrI material. Phonon dispersion curves at different strains revealed this material to be dynamically stable upto 10% strain. The calculated electronic band structure signifies LaBrI to be an indirect band gap (0.51 eV) ferromagnetic material with a magnetic moment of 1.0 $\mu_B$. Band gap with spin-up and spin-down charge carriers can be effectively modulated by biaxial tensile strain and it has a crucial impact on the thermoelectric transport properties. Computed thermoelectric coefficients suggest that this material can have a high power factor with n-type and p-type dopings and low lattice thermal conductivities of ~ $0.93 W/m.K$ that results in a high *ZT* ~ 1.84 at 800K at 4% strain. Thus, computed *ZT* >1 for strained monolayer is comparable to the commercially used TE materials such as $Bi_2Te_3$, oxides, sulphides. Such high value of ZT indicates the potential of this monolayer for thermoelectric applications.

## Acknowledgments

We would like to acknowledge the fellowship and computational resources provided by the Indian Statistical Institute, Kolkata, those have been helpful during this research. Also, we would like to acknowledge the developers of Quantum espresso for providing free open-source codes. RKB acknowledges SERB-DST (EEQ/2022/000325) for the financial support.

# Supplemental Material: Enhanced Thermoelectric Properties of 2D Janus Ferromagnetic LaBrI with Strain-induced Valley Degeneracy

Anuja Kumari,[1, a] Raman D. K,[2] Tanu Choudhary,[3] Raju K Biswas,[3, 4] Abhinav Nag,[5] Santanu K. Maiti,[1, b] Jagdish Kumar[2]

[1]*Physics and Applied Mathematics Unit, Indian Statistical Institute, 203 Barrackpore Trunk Road, Kolkata-700 108, India*
[2]*Department of Computational Sciences, Central University of Punjab, Bhatinda-151 401, India*
[3]*Department of Physics, M S Ramaiah University of Applied Sciences, Bengaluru-560 058, India*
[4]*Department of Physics, North Eastern Regional Institute of Science and Technology, Nirjuli, Arunachal Pradesh-791 109, India*
[5]*Department of Physics, GGDSD college, Rajpur, Palampur-176 061, India*

## Additional data on thermal conductivity:

The convergence of thermal conductivity with q-mesh size is very crucial and presented in Fig. S1. Various q-mesh sizes were considered to compute the lattice thermal conductivity, and a converged set of $20 \times 20 \times 1$ q-mesh was used for further calculations.

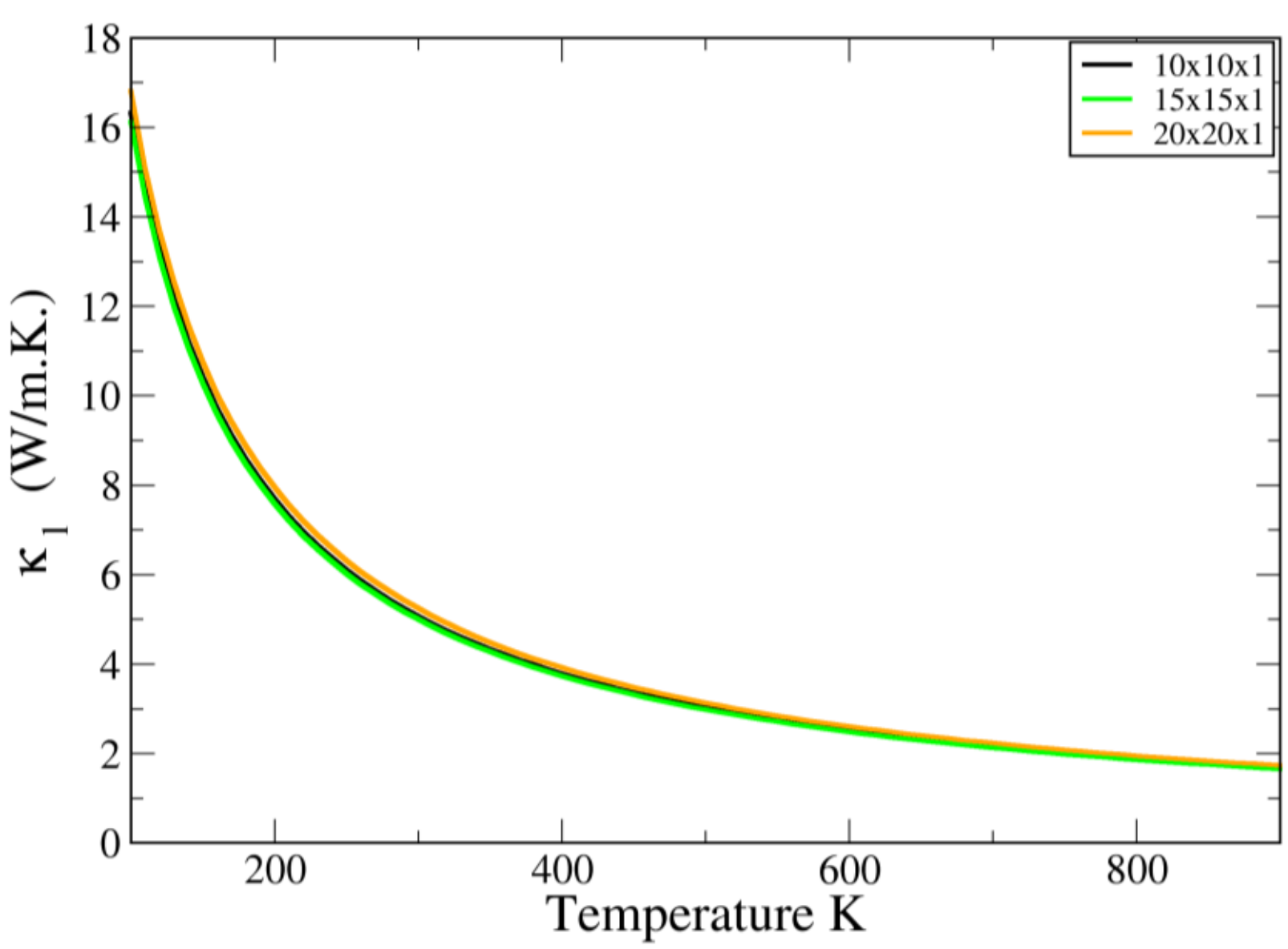

Fig. **S1**: (Color online.) Convergence of thermal conductivity at various *q*-mesh sizes.

## Data for high symmetry path:

The phonon dispersion curves along selected high symmetry K-points ($M - \Gamma - K - M$) were computed with the following coordinates.

The high symmetry $K$-points ($M - \Gamma - K - M$)

| | | | |
|---|---|---|---|
| $M$ | -0.5000000000 | 0.5000000000 | 0.0000000000 |
| $\Gamma$ | -0.0000000000 | 0.0000000000 | 0.0000000000 |
| $K$ | -0.3333333333 | 0.6666666667 | 0.0000000000 |
| $M$ | -0.5000000000 | 0.5000000000 | 0.0000000000 |